\documentclass{article}

\usepackage{arxiv}

\usepackage[utf8]{inputenc} 
\usepackage[T1]{fontenc}    
\usepackage{hyperref}       
\usepackage{url}            
\usepackage{booktabs}       
\usepackage{amsfonts}       
\usepackage{nicefrac}       
\usepackage{microtype}      
\usepackage{lipsum}
\usepackage{cite}
\usepackage{amsmath,amssymb,amsfonts}
\usepackage{algorithmic}
\usepackage{graphicx}
\usepackage{textcomp}
\usepackage{epstopdf}
\usepackage{xcolor}
\usepackage{subfig}
\usepackage{siunitx}
\usepackage{textcomp}
\usepackage{amsmath}

\title{THz graphene W-shaped three-port circulator with dynamical control}

\author{
  Victor Dmitriev\thanks{Use footnote for providing further
    information about author (webpage, alternative
    address)---\emph{not} for acknowledging funding agencies.} \\
 Department of Electrical Engineering\\
  Federal University of Para\\
  66075-900, Belem, Para, Brazil. \\
  \texttt{victor@ufpa.br} \\
   \And
Geraldo Melo\\
  Federal Rural University of Amazonia\\
 66.077-830, Belem, Para, Brazil.\\
    \texttt{geraldo.melo@ufra.edu.br} \\
  \AND
  Wagner Castro\\
  Institute Cyberspace\\
  Federal Rural University of Amazonia\\
 66.077-830, Belem, Para, Brazil.\\
    \texttt{wagner.ormanes@ufra.edu.br} \\
  \And
  Cristiano Braga\\
  Department of Electrical Engineering\\
 Federal University of Para\\
  66075-900, Belem, Para, Brazil. \\
  \texttt{cboliveira@ufpa.br} \\
  \And
  Daimam Zimmer\\
  Department of Electrical Engineering\\
 Federal University of Para\\
  66075-900, Belem, Para, Brazil. \\
  \texttt{zimmer@ufpa.br} \\
}

\begin{document}
\maketitle

\begin{abstract}

We propose  a new  graphene-based THz circulator. It consists of a circular graphene resonator and three graphene nanoribbon waveguides of W-geometry placed on a dilectric substrate. Surface plasmon-polariton waves propagate in the waveguides. The nanoribbon excites  in the resonator  dipole resonance. Nonreciprocity of the device is defined by nonsymmetry of the  conductivity tensor of the magnetized graphene.   The circulator with the central frequency  7.5 THz  has the bandwidth  4.25\% for  isolation -15 dB,   insertion loss  -2.5 dB. Applied DC magnetic field  is 0.56T and Fermi energy of graphene  $\epsilon_F=0.15$eV.   The Fermi energy  allows one to control dynamically the circulator responses.
\end{abstract}


\section{INTRODUCTION}

\label{intro}

	Three-port circulator is one of the most popular nonreciprocal devices in microwave technology due to its excellent electrical characteristics,  simple structure and versatility in applications. Typically, such a device has a three-fold rotational symmetry with input port, output port and one decoupled port \cite {Helszain1}. The ports are usually microstrip  lines or rectangular metal waveguides. The theory of stripline and waveguide microwave circulators is well developed \cite {Helszain2}. Circulators of different types for optical systems are also described in literature (see, for example, \cite {circulators, circulators1}). The main function of circulators is  protection of a source of signal from unwanted reflections in the circuit, directing the reflected signal to the decoupled port where a matched load is connected. They are used also as switches \cite {switch1, switch2}, as wavelength multiplexers and demultiplexers \cite {mult1,mult2}, in dispersion compensators \cite {compensador}, in sensors \cite {sensor} and reflectometry \cite {reflectometria}.

The physical effects utilized to produce the circulation are nonreciprocal phase shift, Faraday effect, nonreciprocal TE - TM mode conversion in waveguides, edge-guided mode regime \cite {Santis} and resonances of two counterrotating modes with different frequencies in circular resonators \cite{Helszain2}. 

At microwaves, the circulators  are usually designed by using ferrites described by nonsymmetrical  permeability tensor $[\mu]$. At higher frequencies, the use of ferrites in  nonreciprocal devices  is limited because of their relatively high losses and limited saturation magnetisation, i.e. weak gyromagnetic effect \cite {ferrite1, ferrite2}. For millimeter wave region, the semiconductor and  2D electron gas  materials \cite {semicondutor}  with nonsymmetrical permittivity $[\varepsilon]$ tensor have been  suggested to overcome some  drawbacks of ferrite materials. 

There are several proposals of the circulators for photonic crystal technology which can be used in THz, infrared and optical regions \cite {circulatorTHz, Magneto, Optical1}.	An idea of nonreciprocal components and, in particular, circulators, where a magnetic system is not required,  was  suggested recently in \cite {Estep} and \cite {Negar}. In such devices the circulation effect is achieved due to spatial-temporal permittivity modulation or staggered commutation. However, from the point of view of working parameters these types of circulators still can not  compete  with the traditional ones.

Nonreciprocal properties of graphene under DC magnetic field have been demonstrated experimentally in Faraday rotation scheme \cite {Faraday}. Recently, it was shown that magnetized graphene due to nonsymmetry of the conductivity tensor $[\sigma]$ can also be used for circulator design in THz and infrared regions.

In this paper, we suggest a new type of graphene-based circulator for the THz region with W-geometry. This device has a very simple and compact structure with good features. The design of the W-circulator was previously suggested in photonic crystal \cite {W-format}.
	
The analytical description of the circulator in terms of the scattering matrix based on temporal couple mode theory and magnetic group theory, which is developed in this work, can serve for modelling complex THz circuits and systems  as it is made in microwave circuitry. Extensive numerical simulations demonstrate a good correspondence with analytical results.
\section{PROBLEM DESCRIPTION}
\begin{figure}[ht]
\centering
{\includegraphics[scale=1]{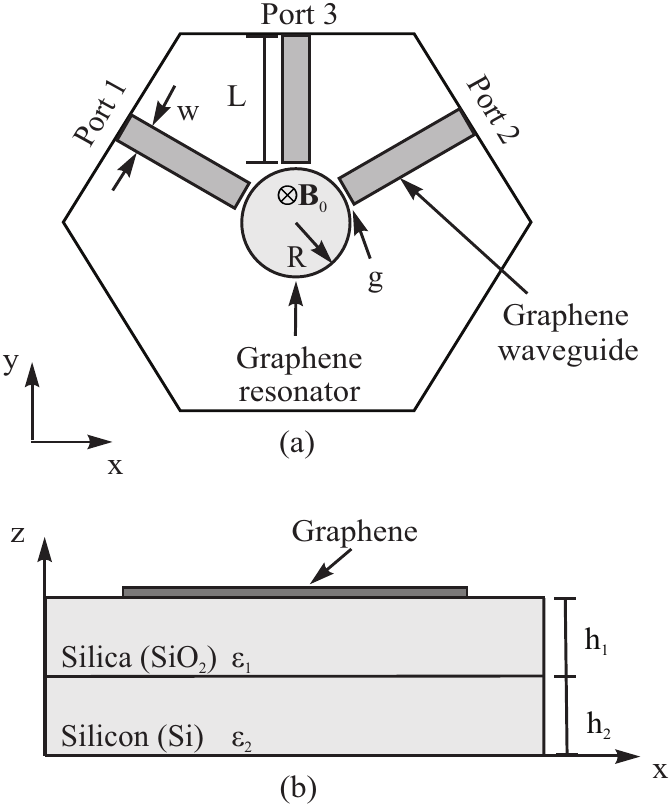}}
\caption{Geometry of W-circulator: a) top view, b) side view, {\bf B}$_0$ is bias DC magnetic field.}
\label{fig:figure1}
\end{figure}

A schematic representation of the three-port W-circulator is shown in Fig. \ref{fig:figure1}. This circulator consists of a circular graphene resonator magnetized by a DC magnetic field normal to graphene layer and three waveguides with angular distance of 60$^{\circ}$ between them. The graphene elements are placed on a two-layer dielectric substrate. One of the layers is silica $(SiO_{2})$ and the other one is silicon $(Si)$.

\begin{figure}[ht]
\centering
{\includegraphics[scale=1]{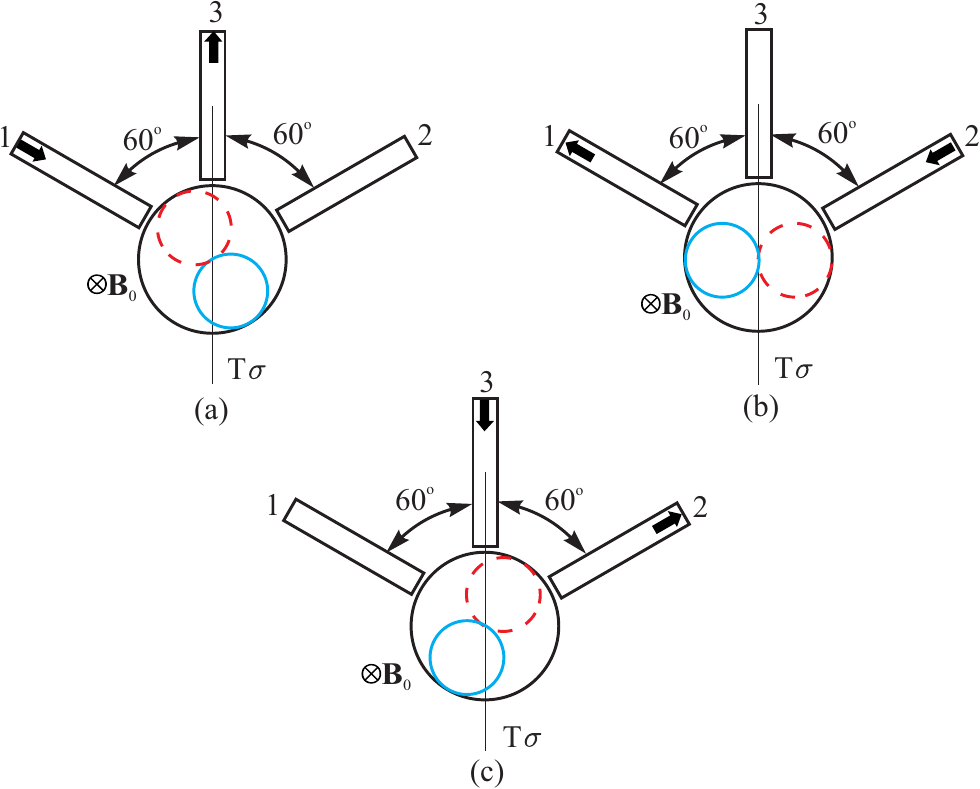}}
\caption{Excitation schemes of W-circulator with dipole mode (dotted and continuous circles) of resonator. Direction of circulation  a) $1\rightarrow3$, b) $2\rightarrow1$, c) $3\rightarrow2$ .}
\label{fig:figure2}
\end{figure}

In  Fig. \ref{fig:figure2}a, \ref{fig:figure2}b and \ref{fig:figure2}c, we demonstrate the physical principle of functioning of circulator considering excitation of different ports. When the signal is injected in port 1, the standing dipole mode has its node in port 2. Therefore, this corresponds to the transmission from port 1 to port 3. The transmission from port 2 to port 1 and from port 3 to port 2 can be explained analogously.

Using group theoretical description of the W-circulator \cite {W-format} and considering the antiplane of symmetry $T\sigma$ (Fig. \ref{fig:figure2}),   the   scattering matrix can be written in the following form:

\begin{equation}
\label{eq:sig01}
	 S=\left(\begin{array}{ccc}
	 S_{11} & S_{12} & S_{13} \\ 
	 S_{21} & S_{11} & S_{23} \\ 
     S_{23} & S_{13} & S_{33}\end{array} \right) 
\end{equation}
where $S_{11} = S_{22}$, $S_{13} = S_{32}$ and $S_{23} = S_{31}$ are due to the antiplane of symmetry $T\sigma$. $S_{11}$ is the reflection coefficient at port 1, $S_{21}$ is the transmission coefficient from port 1 to port 2, $S_{31}$ is the transmission coefficient from port 1 to port 3, and so on. The per-cent fractional bandwidth of the device is defined as  $BW = (f_2 - f_1)/f_0 \times 100\%$, where $f_2$ and $f_1$  are the upper and  lower frequency defined by the transmission level below -2.5 dB and the isolation level of the isolation port higher than -15 dB, $f_0$ is the  central frequency of operation.	

	The device has the following dimensions: the length and width of the waveguides are $L = 900$ nm and $w = 120$ nm, respectively, the resonator radius $R = 320$ nm, the gap between the waveguides and the resonator is $g = 5$ nm. The dielectric layers of $SiO_{2}$ and $Si$ have the thicknesses $h_{1} = h_{2}= $ 2500 nm, with relative permittivity $\varepsilon_1 = $ 2.09 and  $\varepsilon_2 = $ 11.9, respectively.
	
	Our task is to analyze the properties of the circulator using the temporal coupled mode theory and numerical methods.
\section{CONDUCTIVITY TENSOR}
Surface conductivity of the magnetized graphene depends on radian frequency $\omega$, Fermi energy $\epsilon_{F}$, which is a function of electric biasing $E_{0}$, and also on the phenomenological scattering rate, defined as $\Gamma = 1/\tau$, ($\tau$ is the relaxation time of graphene) and the cyclotron frequency $\omega_B = {eB_0v_{F}^{2}}/{\epsilon_F}$, where $e$ is the electron charge, $v_F$ is the Fermi velocity, $B_0$ is DC  magnetic field. 

\begin{figure}[ht]
\centering
{\includegraphics[scale=1]{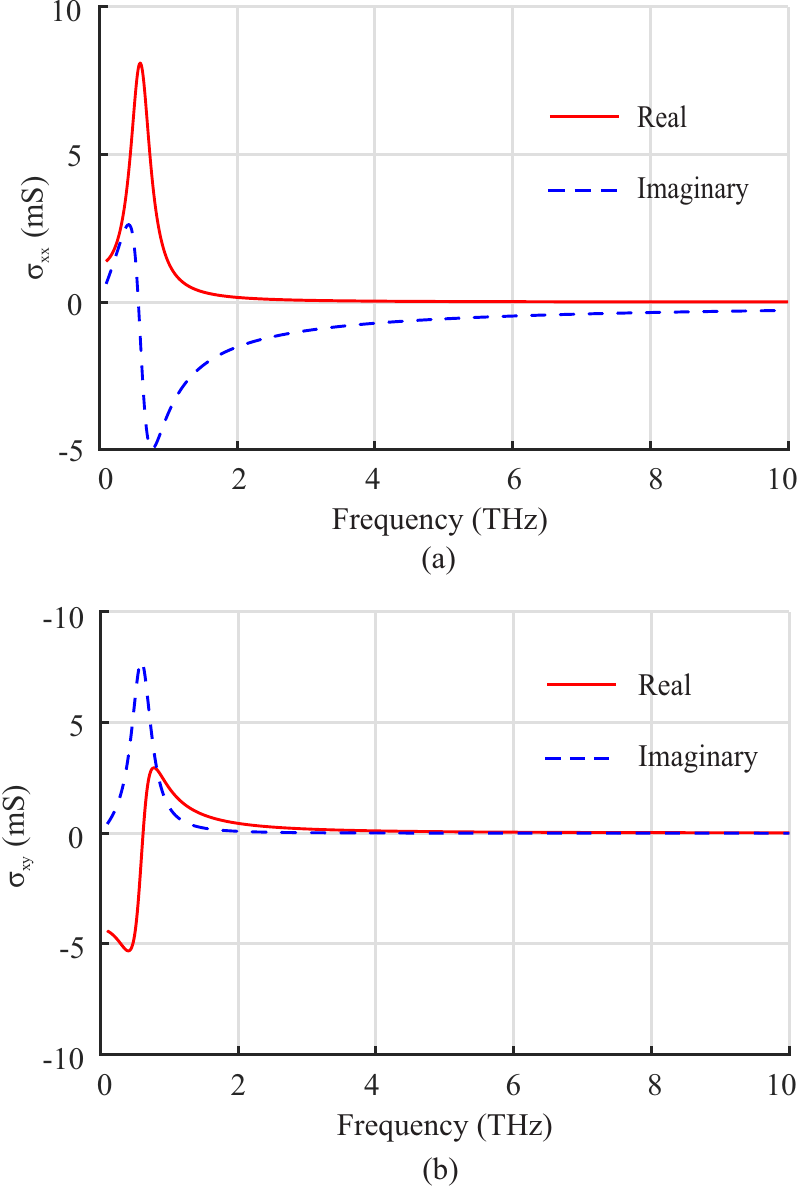}}
\caption{Real and imaginary parts of components $\sigma_{xx}$ ans $\sigma_{xy}$ of conductivity tensor of graphene, $B_{0}=0.56$ T, $\epsilon_{F} = 0.15$ eV.}
\label{fig:figure3}
\end{figure}

The tensor of the graphene conductivity \cite{sigma} is:
\begin{equation}
\label{eq:sig1}
	[\sigma_s(\omega, \epsilon_{F}({E}_{0}), \Gamma, \omega_{B})]=\left[\begin{array}{cc}
	 \sigma_{xx} & -\sigma_{xy} \\ 
	 \sigma_{xy} & \sigma_{xx}\\ 
    \end{array} \right],
\end{equation}
where $\sigma_{xx}=\sigma_{yy}$ and  $\sigma_{yx}=-\sigma_{xy}$ are longitudinal (diagonal) and transverse (off-diagonal) parts, respectively. For the regime of  intraband transitions in the THz region, they  can be defined by the Drude formula \cite{inter}:
\begin{eqnarray}
\label{eq:sig2}
\sigma_{xx}=\frac{2D}{\pi} \frac{1/\tau-i\omega}{ \omega_{B}^{2}-( \omega + i / \tau ) ^2}, \\
\label{eq:sig3a}
\sigma_{xy}=-\frac{2D}{\pi} \frac{\omega_B}{ \omega_{B}^{2}-( \omega+i / \tau )^2}.
\end{eqnarray}
where the Drude weight is given by $D=2\sigma_0 \epsilon _F / \hbar$,  $\hbar$ is the reduced Planck's constant and $\sigma_0=e^2/(4\hbar)$ is the minimum conductivity of graphene.
Fig. \ref{fig:figure3}a and Fig. \ref{fig:figure3}b show the frequency dependence of  the real part and of the imaginary one of the tensor components of the graphene electrical conductivity. It can be observed that the cyclotron resonance frequency $\omega_{B}$ of the graphene is around 0.6 THz and that the imaginary part of the conductivity $\sigma_{xx}$ becomes negative above this frequency, which is necessary condition for the graphene device to support transverse magnetic (TM)
surface plasmon polariton (SPP) waves. On the other hand, the real part of the conductivity, responsible for the graphene losses, increases rapidly in the vicinity of this frequency. So, to reduce losses in the circulator, the  operating frequency of the device must be sufficiently higher than the cyclotron resonant frequency of the graphene.
\section{GRAPHENE NANORIBBON WAVEGUIDE}
The numerical simulations were performed using the commercial software COMSOL Multiphysics version 5.2a \cite{comsolSite}, which is based on the Finite Element Method. Graphene  has a very small thickness which can not be inserted directly in the modelling by the COMSOL software. To overcome this difficulty, we consider a monolayer with a finite thickness $\Delta$ and the conductivity tensor given by $[\sigma_v] = [\sigma_s]/\Delta$, where $[\sigma_s]$ is the tensor surface conductivity of graphene (\ref{eq:sig1}) given by components (\ref{eq:sig2}) and (\ref{eq:sig3a}). In our numerical simulations we assume $\Delta = 1$ nm. The artificial parameter $\Delta$ is used here only for calculation purposes \cite{Transformation} and numerical calculations were made for the following data: $\epsilon_{F} = 0.15$ eV, $B_{0}=0.56$ T and $\tau=0.9$ ps \cite {tau}.

\begin{figure}[ht]
\centering
{\includegraphics[scale=1]{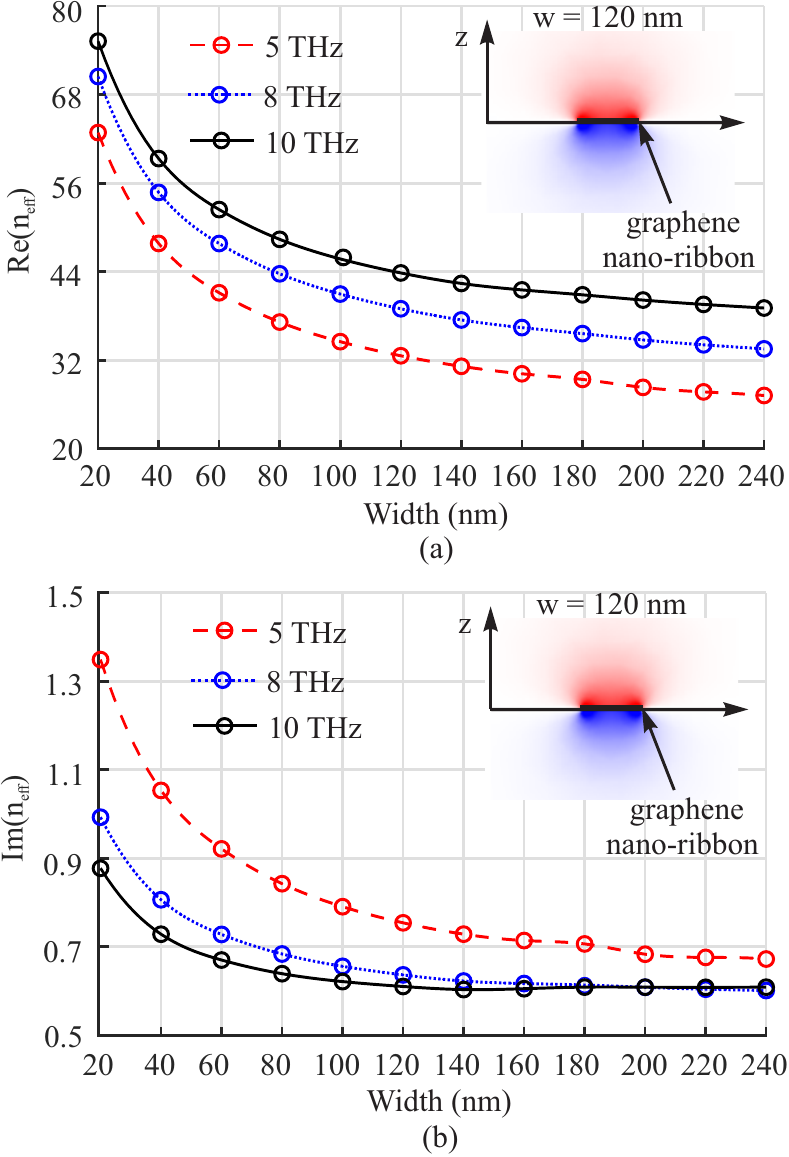}}
\caption{Width dependence mode of effective index for  5 THz, 8 THz, 10 THz. Insert is the $E_z$  profile of guided mode in the graphene strip.}
\label{fig:figure4}
\end{figure}

	In literature, the two main types of modes guided in graphene ribbons are discussed. One of them is concentrated around the center of the nanoribbon and other one at nanoribbon edges, both are known as surface plasmon polariton (SPP) modes \cite{Nikitin2011, HE2013}. For our purposes, we  choose the first mode which is shown in the insert of Fig. \ref{fig:figure4}. In this figure, the dependence of the effective refractive index with respect to the width of the geraphene nanoribbon is presented.
\section{RESONATOR RADIUS}
	The radius $R$ of the resonator for the lowest dipole mode can be estimated approximately from the condition $2\pi R=\lambda_{spp}$, where $\lambda_{spp}$ is the wavelength of the TM SPP mode. It is well known that the infinite graphene placed on an air-dielectric interface supports transverse-magnetic TM SPP waves with a dispersion relation given by: 
\begin{equation}
\beta_{spp}=\dfrac{(1+\varepsilon_1)(\omega\hbar)^2}{4\alpha\epsilon_F\hbar c}\biggl(1-\dfrac {\omega_{B}^2}{\omega^2}\biggr),
\label{eq:sig3}
\end{equation} 
where $\alpha=e^2/(4\pi \varepsilon_0 \hbar c) \approx 0.007$ is the fine-structure constant \cite {Livro} and $c \approx 3\times10^{8}$ m/s is the light velocity. Using the relations $R=\lambda_{spp}/2\pi$ and $\beta_{spp}=2\pi/\lambda_{spp}$,  the radius of the resonator can be defined as follows:  
\begin{equation}
\label{eq:sig4}
R\approx A\dfrac{\epsilon_F}{(1+\varepsilon_1)(\omega^2-\omega_{B}^2)},
\end{equation}
where  $A = 8.3\times10^{40} (kg.m)^{-1}$, $\epsilon_{F}$ is Fermi energy given in $(J)$. As follows from (\ref{eq:sig4}),  the radius of the resonator $R$ in $(m)$ is defined by the working frequency of the circulator $\omega$,  cyclotron frequency of graphene $\omega_{B}$ given in (Hz),   $\epsilon_F$ and  by the permittivity of the substrate $\varepsilon_1$.

\section{SIMULATION RESULTS}
\subsection{MAGNETIZED RESONATOR}\label{AA}
\begin{figure}[h!]
	\centerline{
		\includegraphics[scale=1]{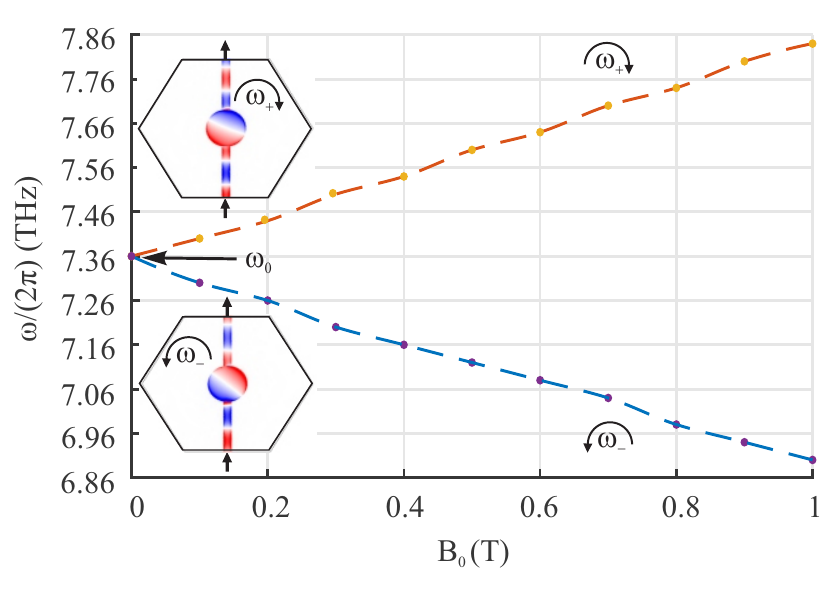}}
	\caption{Magnetic DC field dependence of rotating mode resonances, $\epsilon_F = 0.15$ eV. Insert: resonator feeding scheme for calculus of resonances.
	}
	\label{figure5}
\end{figure}

	Plasmonic waves are guided by graphene nanoribbons (see  Fig \ref{fig:figure4}). They excite dipole  plasmonic resonance in the circular resonant graphene element. The SPP wave of the input waveguide excites in the circular resonator two degenerate clockwise $\omega_{+}$ and anticlockwise $\omega_{-}$ rotating modes. In Fig \ref{figure5}, the DC magnetic field $B_0$ dependence of the frequencies of the two rotating modes is shown. For the calculus, we use  the model of only one exciting waveguide, one output waveguide and the resonant cavity (see insert in Fig \ref{figure5}). The splitting of the  modes $\omega_{+}$ and $\omega_{-}$  increases with enlargement of $B_0$ almost linearly. The value of   ${B}_0$  which provides at the central frequency of the circulator band the maximum of isolation, will be considered in  the following discussion as an optimal DC magnetic field. 
	
\begin{figure}[h!]
	\centerline{
		\includegraphics[scale=1]{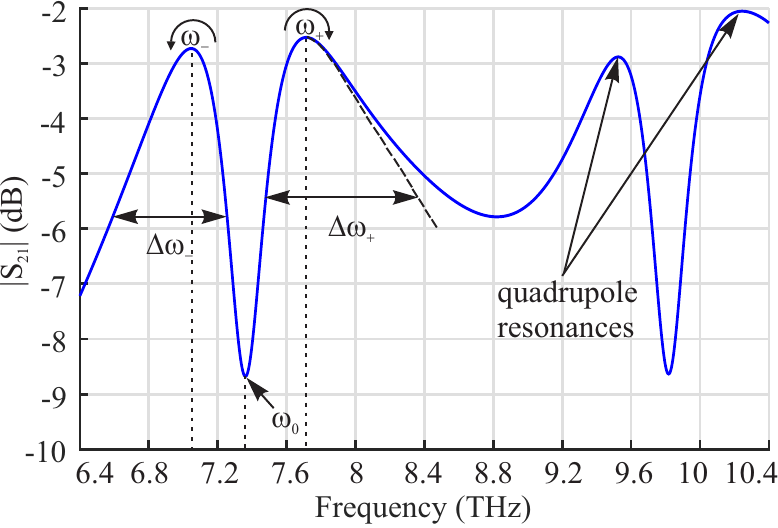}}
	\caption{Frequency responses of resonant modes $\omega_{+}$ and $\omega_{-}$ for the feeding scheme shown in insert of Fig. \ref{figure5}. $B_{0}$ = 0.56 T, $\epsilon_F = 0.15$ eV.
	}
	\label{figure6}
\end{figure}

	In  Fig. \ref{figure6} one can observe peaks of the transmission coefficient in 7.1 THz and 7.7 THz, corresponding to the dipole rotating modes with frequencies $\omega_{-}$ and $\omega_{+}$ and the quadrupole rotating modes in the interval from 9.5 THz to 10.2 THz. To calculate the Q-factor for the $\omega_{+}$ mode, we draw a line separating modes $\omega_{+}$ of the dipole and quadrupole resonances, because as can be seen, there is a superposition of these modes in the frequency range between them. 

\subsection{CIRCULATOR RESPONSES}
\begin{figure}[h!]
	\centerline{
		\includegraphics[scale=1]{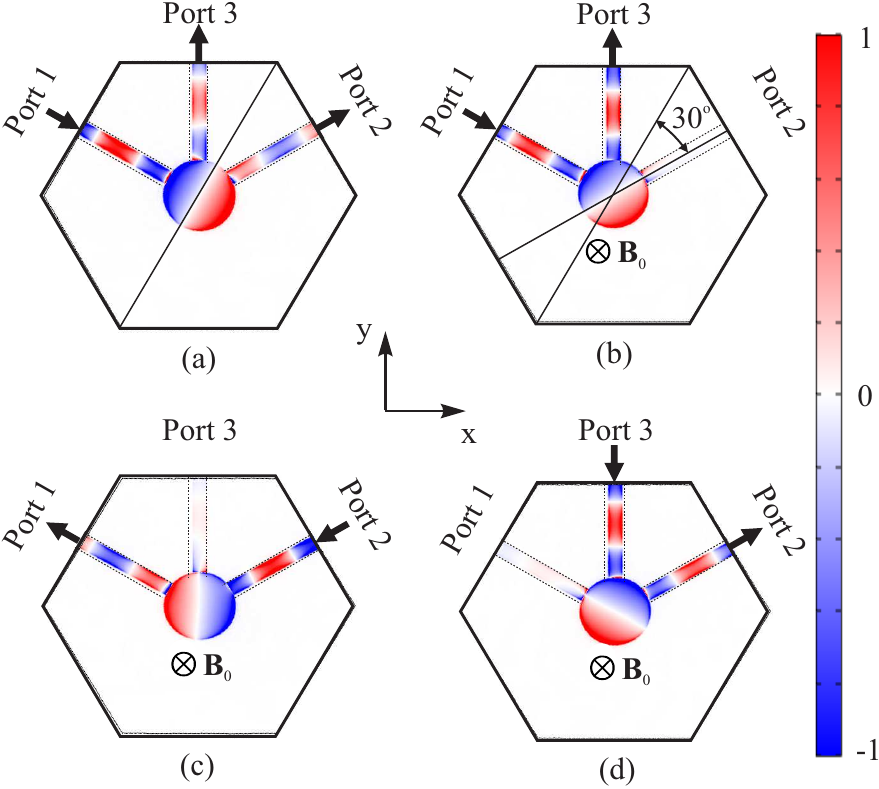}}
	\caption{$E_z$ field profile for a) nonmagnetized case, and also for magnetized cases  b) for transmission (1$\rightarrow$3), c)  (2$\rightarrow$1) and d) (3$\rightarrow$2).  
}
	\label{figure7}
\end{figure}

\begin{figure}[h!]
	\centerline{
		\includegraphics[scale=1]{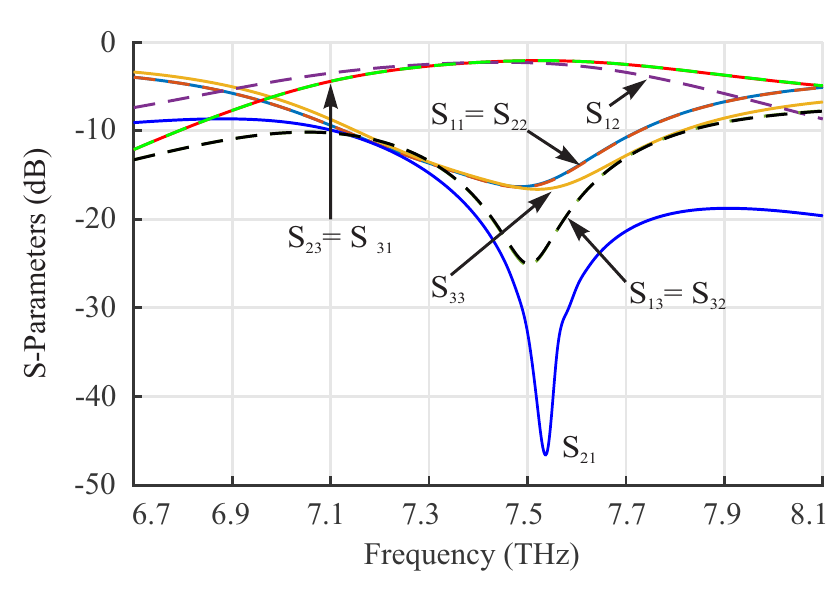}}
	\caption{Frequency responses of circulator, $R$ = 320 nm, $g$ = 5 nm, $w$ = 120 nm, $B_{0} = 0.56$ T, $\epsilon_F = 0.15$ eV, excitation at port 1, 2 and 3.
}
	\label{figure8}
\end{figure}

For the case of the three-port without magnetization shown  in Fig \ref{figure7}a, the sum of the counter-rotating modes produces a standing dipole, which leads to a transmission in both output ports. Magnetization by DC magnetic field breaks the degeneracy of the rotating modes and makes the field pattern of the standing wave to rotate by 30$^{\circ}$ aligning thus the node of the dipole to port 2 and, therefore, isolating  this port and transmitting the electromagnetic wave to port 3. Thus, if the signal is injected into  port 1, it will be transmitted to port 3 with port 2 isolated. If the signal is injected in port 2, it is transmitted to port 1, isolating  port 3. The same is true in the case of incidence in  port 3: the signal is transmitted to port 2 and port 1 is isolated, as shown in Fig \ref{figure7}b, \ref{figure7}c, \ref{figure7}d, respectively. 
	The frequency characteristics for the cases presented in Fig. \ref{figure7}b, \ref{figure7}c and \ref{figure7}d are shown in Fig. \ref{figure8}. For the case shown in \ref{figure7}c, the central frequency of operation is  7.5 THz, the device has the transmission coefficient of - 2 dB and isolation of -24.3 dB, presenting the bandwidth of 4.25\% at the isolation level  of -15 dB.	

Fig. \ref{figure8} shows a comparison of the transmission, isolation and reflection curves of circulator, witch are in accordance with the symmetry analysis (see matrix (\ref{eq:sig01})), where the isolations $S_{13}$ and $S_{32}$ are coincident, as well as the transmissions $S_{23}$ and $S_{31}$ and also the reflections $S_{11}$ and $S_{22}$. The peak of $S_{21}$ is higher and the curve $S_{21}$ presents a larger bandwidth with respect to the other two characteristics.
\section{TEMPORAL COUPLED MODE THEORY OF CIRCULATOR}\label{AA}
	The temporal coupled mode theory (TCMT) \cite {Joannopolus} is based on very general assumptions such as weak coupling of the resonator with waveguides, linearity of the system, Time-reversal symmetry and energy conservation. It has been used in \cite {Y-circulator} to calculate  characteristics of the Y-circulator. In the following, we use the theory developed   in \cite {w-circulator} for W-circulator. 
	
	In the following,  we suppose, in the same way as in \cite {Y-circulator}, that  for the broken Time reversal symmetry, nonreciprocity of the device is defined only by the effect of the frequency splitting of the counter-rotating $\omega_{+}$ and $\omega_{-}$ of the resonator and the coupling coefficients between the waveguides and the resonator are reciprocal. The scattering 	matrix elements are defined as follows \cite {w-circulator}:
\begin{equation}
S_{11}=S_{22}=S_{33}= -1 +\frac{2}{3}\biggl(\frac{\gamma_{+w}}{\gamma_{+}+i(\omega-\omega_{+})}+
\frac{\gamma_{-w}}{\gamma_{-}+i(\omega-\omega_{-})}\biggr),
\label{eq17}
\end{equation}
\begin{equation} 
S_{12}=\frac{2}{3}e^{-i2\varphi/3}\biggl(\frac{e^{i2\pi/3}\gamma_{+w}}{\gamma_{+}+i(\omega-\omega_{+})}+
\frac{e^{-i2\pi/3}\gamma_{-w}}{\gamma_{-}+i(\omega-\omega_{-})}\biggr),
\label{eq18}
\end{equation}
\begin{equation} 
S_{32}=S_{13}=\frac{2}{3}e^{-i2\varphi/3}\biggl(\frac{e^{-i2\pi/3}\gamma_{+w}}{\gamma_{+}+i(\omega-\omega_{+})}+
\frac{e^{i2\pi/3}\gamma_{-w}}{\gamma_{-}+i(\omega-\omega_{-})}\biggr),
\label{eq19}
\end{equation}
\begin{equation} 
S_{21}=\frac{2}{3}e^{i2\varphi/3}\biggl(\frac{e^{-i2\pi/3}\gamma_{+w}}{\gamma_{+}+i(\omega-\omega_{+})}+
\frac{e^{i2\pi/3}\gamma_{-w}}{\gamma_{-}+i(\omega-\omega_{-})}\biggr),
\label{eq20}
\end{equation}
\begin{equation} 
S_{31}=S_{23}=\frac{2}{3}e^{i2\varphi/3}\biggl(\frac{e^{i2\pi/3}\gamma_{+w}}{\gamma_{+}+i(\omega-\omega_{+})}+
\frac{e^{-i2\pi/3}\gamma_{-w}}{\gamma_{-}+i(\omega-\omega_{-})}\biggr),
\label{eq21}
\end{equation}
\linebreak
where $\varphi = \varphi_{21} - \varphi_{32}$, $\varphi_{21}$ and $\varphi_{32}$ are phases of the elements $S_{21}$ and $S_{32}$, respectively. $\gamma_{+}=\gamma_{+w}+\gamma_{+i}$ and $\gamma_{-}=\gamma_{-w}+\gamma_{-i}$ are decay rates. Notice that $\gamma_{+w}$ and $\gamma_{-w}$ are decay rates due to coupling of the resonator with waveguides, and  $\gamma_{+i}$ and $\gamma_{-i}$ due to internal losses of the resonator. 

The resonance frequencies $\omega_{+}$, $\omega_{-}$ and the decay rates $\gamma_{+}$ and $\gamma_{-}$ were obtained from computational simulations for the scheme in the insert of Fig. \ref{figure5}. More specifically, the parameters of the TCMT $\omega_{+}$, $\omega_{-}$, $\gamma_{+}$ and $\gamma_{-}$ were obtained from the Comsol simulations shown in Fig. \ref{figure6}. The decay rate $\gamma_{+(_-)}$ is related to the quality factor $Q_{+(_-)}$, as $Q_{+(_-)} = (3/2)\omega_{+(_-)}/2\gamma_{+(_-)}$. To compensate  the difference in the number of ports of the devices (the three-port circulator and the two-port of Fig. \ref{figure5}), we multiplied  the Q-factor by the constant 3/2. The obtained parameters  are $\omega_{+}=48.38\times10^{12}$ rad/s,  $\omega_{-}=45.24\times10^{12}$ rad/s, $\gamma_{+}\cong3.4\times10^{12}$ 1/s, $\gamma_{-}\cong3.1\times10^{12}$ 1/s. The parameters corresponding to internal losses of the resonator $\gamma_{+i}\cong0.9\times10^{12}$ 1/s and $\gamma_{-i}\cong0.25\times10^{12}$ 1/s, were used as fitting parameters to adjust the TCMT and the Comsol curves. The decay rates due to coupling of the resonator with waveguides are $\gamma_{+w}\cong2.5\times10^{12}$ 1/s and $\gamma_{-w}\cong2.75\times10^{12}$ 1/s.  
\begin{figure}[h!]
	\centerline{
		\includegraphics[scale=1]{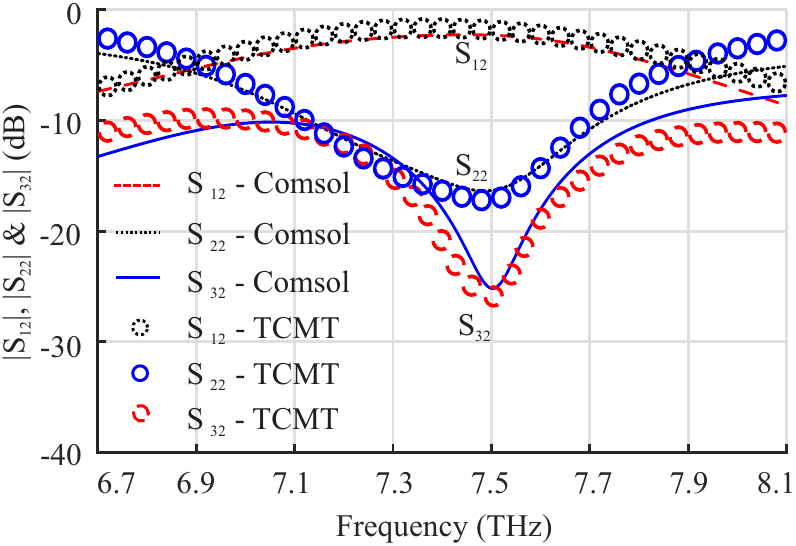}}
	\caption{Frequency responses of circulator simulated by Comsol (continuous lines) and by TCMT (circles), $R$ = 320 nm, $g$ = 5 nm, $w$ = 120 nm, $B_{0} = 0.56$ T, $\epsilon_F = 0.15$ eV, excitation at port 2.
	}
	\label{figure9}
\end{figure}

The theoretical results obtained by equations (\ref{eq17}) - (\ref{eq21}) and by software Comsol are  shown in Fig. \ref{figure9}. We present only the results for the case  of excitation applied at port 2, since similar results can be obtained for the excitation of ports 1 and 3. It can be seen that there is a good accordance between the two methods.

	Similar to the photonic crystal W-circulator \cite {w-circulator}, the ideal circulator response with $S_{12} = 1$ and $S_{32} = 0$ ($\gamma_{+i} = \gamma_{-i} = 0$) can be obtained at the central frequency of operation $\omega_{c}$. Bandwidth $BW$ for ideal circulator at the level of isolation -15 dB can be estimated by (\ref{eq19}) and is defined by
\begin{equation}
\label{eq:27}
BW =0.65\dfrac{(\omega_{+}-\omega_{-})}{\omega_{c}} 100\%.             
\end{equation}
\section{PARAMETRIC ANALYSIS OF CIRCULATOR}
\subsection{Varying the Radius of Resonator}
\begin{figure}[h!]
	\centerline{
		\includegraphics[scale=1]{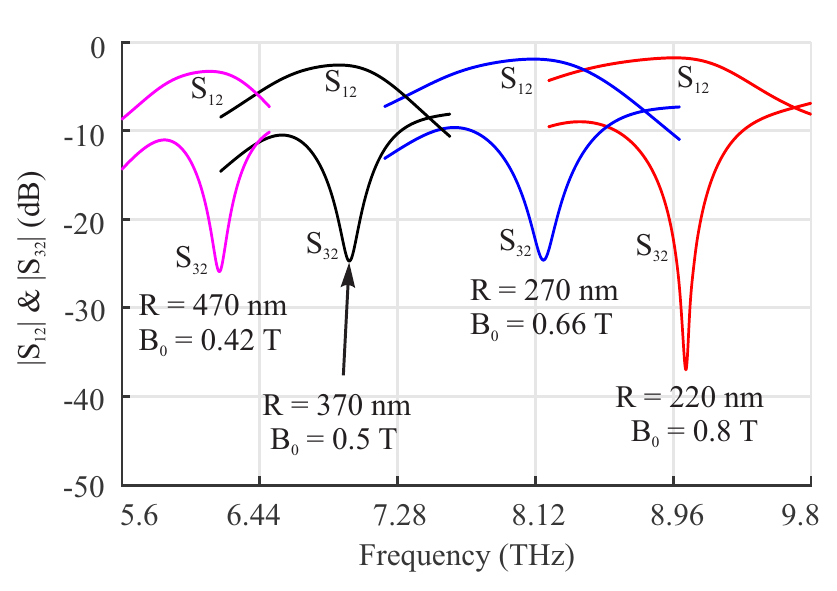}}
	\caption{Frequency characteristics of circulator with  different radii $R$, $g$ = 5 nm, $w$ = 120 nm, $\epsilon_{F} = 0.15$ eV, excitation at port 2.
	}
	\label{figure10}
\end{figure}

\begin{figure}[h!]
	\centerline{
		\includegraphics[scale=1]{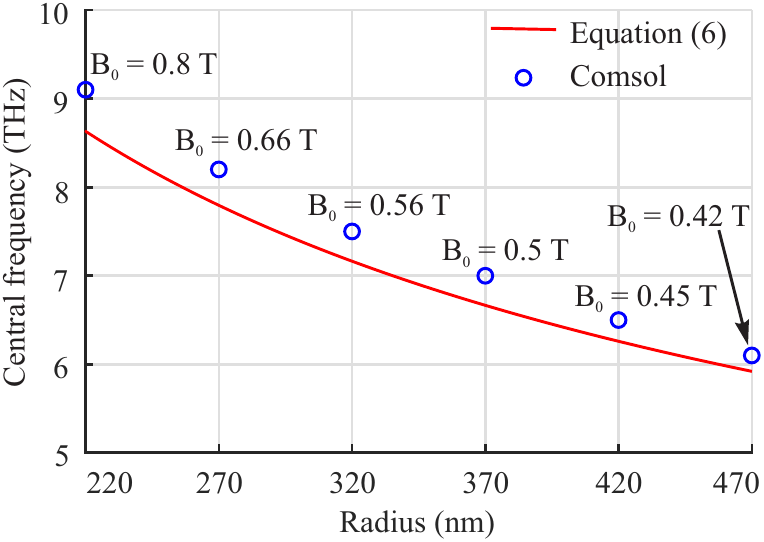}}
	\caption{Central frequency of circulator versus radii $R$, $g$ = 5 nm, $w$ = 120 nm, $\epsilon_{F} = 0.15$ eV. Numbers on the curve are DC magnetic field $B_0$, excitation at port 2.
	}
	\label{figure11}
\end{figure}

\begin{figure}[h!]
	\centerline{
		\includegraphics[scale=1]{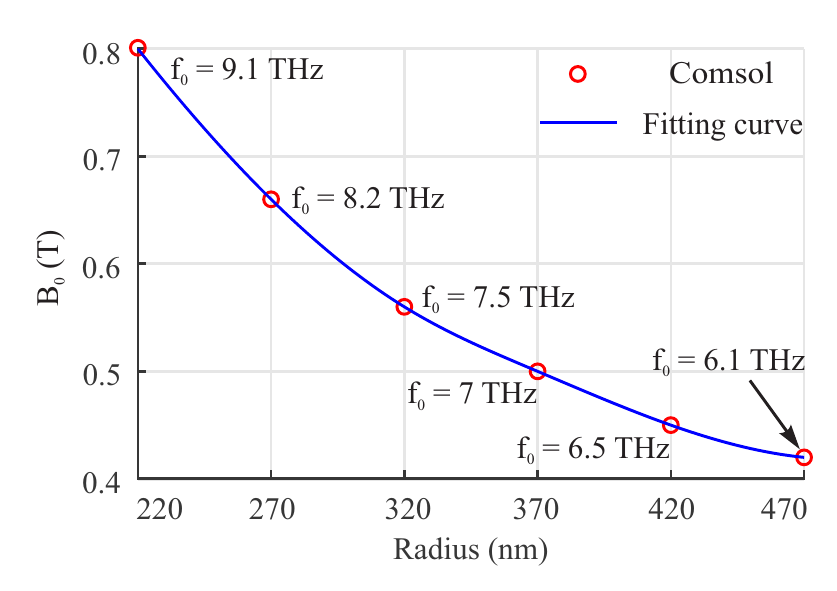}}
	\caption{Optimal magnetic field of circulator for different radii $R$, $g$ = 5 nm, $w$ = 120 nm, $\epsilon_{F} = 0.15$ eV. Numbers on the curve are central frequency ${f}_{0}$, excitation at port 2.
	}
	\label{figure12}
\end{figure}

	The influence of the radius of the resonator on the resonance frequencies  was investigated. The radius was changed from 220 nm to 470 nm and the other parameters as Fermi energy, width and gap were fixed at $\epsilon_F = 0.15$ eV, $w = 120$ nm, $g$ = 5 nm, respectively. Fig. \ref{figure10} shows the frequency responses of the circulator corresponding to the radii of 220 nm, 270 nm, 370 nm and 470 nm. As can be seen from Fig. \ref{figure11}, the central frequency of the circulator is increased with decreasing the radius $R$ of the resonator. One can observe a systematic error of approximately 8\% between the simulated results and those obtained by equation (\ref{eq:sig4}). This difference can be compensated including the multiplicative factor 1.08 in equation (\ref{eq:sig4}).  One can also see in Fig. \ref{figure12}, that the optimal DC magnetic field $B_0$ decreases with increase of the resonator radius.  
	\subsection{Influence of Waveguide  Width}
\begin{figure}[h!]
	\centerline{
		\includegraphics[scale=1]{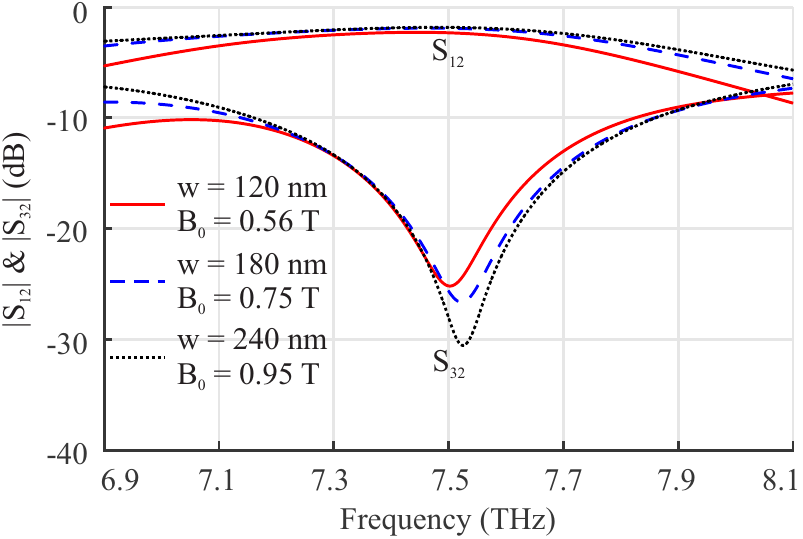}}
	\caption{Frequency responses of circulator for different widths $w$, $g$ = 5 nm, $R = 320$ nm, $\epsilon_F = 0.15$ eV, excitation at port 2.
	}
	\label{figure13}
\end{figure}

	The influence of the width $w$ of the graphene waveguide on the resonance frequency and the insertion losses of the device was investigated as well. Varying the width from 120 nm to 240 nm we kept the other parameters  fixed, i.e. $\epsilon_F = 0.15$ eV, $R$ = 320 nm and $g$ = 5 nm. The frequency responses of the circulator are shown in Fig. \ref{figure13}. There is a shift of the resonance frequencies to higher values with increasing the width of the waveguide.      	
\subsection{Gap Dependence of Circulator  Characteristics}
\begin{figure}[h!]
	\centerline{
		\includegraphics[scale=1]{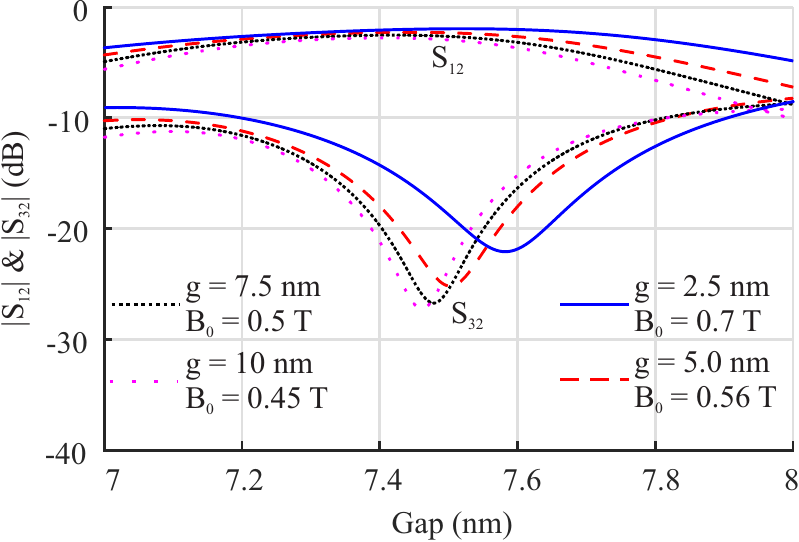}}
	\caption{Frequency responses of circulator for different values of gap $g$, $w$ = 120 nm, $R = 320$ nm, $\epsilon_F = 0.15$ eV, excitation at port 2.
	}
	\label{figure14}
\end{figure}

\begin{figure}[h!]
	\centerline{
		\includegraphics[scale=1]{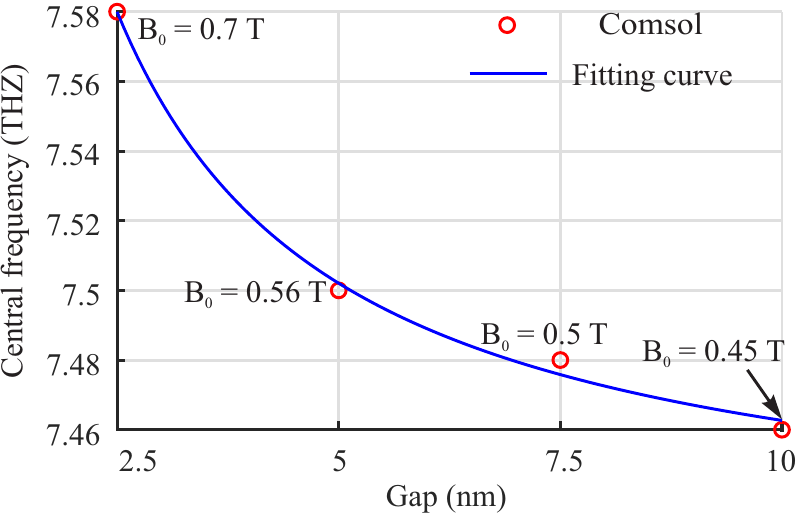}}
	\caption{Central frequency of circulator for different values of gap $g$, $w$ = 120 nm, $R = 320$ nm, $\epsilon_F = 0.15$ eV. Numbers on the curve are DC magnetic field $B_0$, excitation at port 2.
	}
	\label{figure15}
\end{figure}

\begin{figure}[h!]
	\centerline{
		\includegraphics[scale=1]{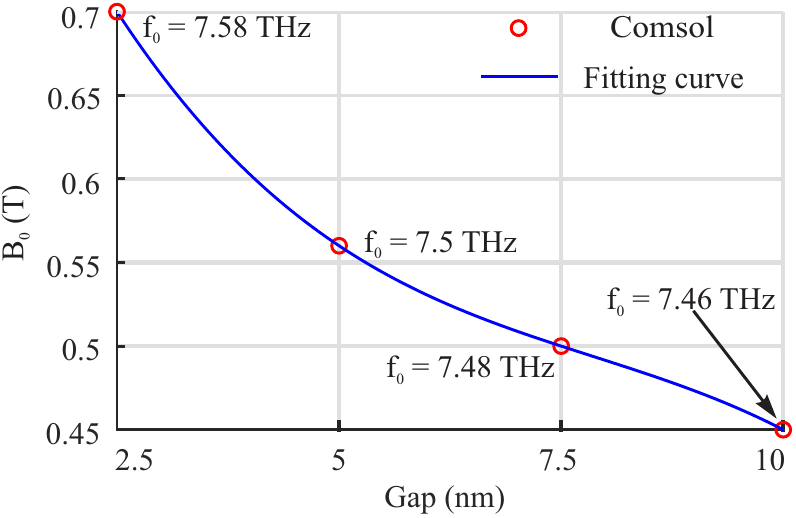}}
	\caption{Gap $g$ dependence of optimal magnetic field of circulator, $w$ = 120 nm, $R = 320$ nm, $\epsilon_F = 0.15$ eV. Numbers on the curve are central frequency ${f}_{0}$, excitation at port 2.
	}
	\label{figure16}
\end{figure}

\begin{figure}[h!]
	\centerline{
		\includegraphics[scale=1]{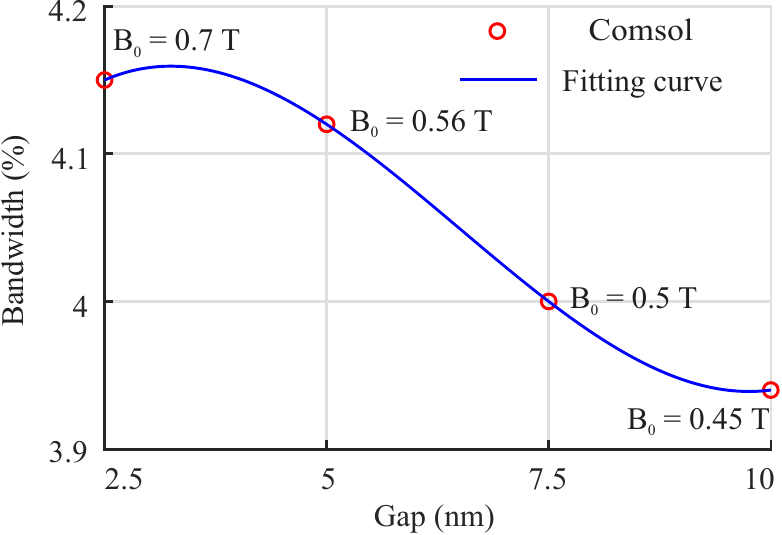}}
	\caption{Bandwidth of circulator for different values of gap $g$; $w$ = 120 nm, $R = 320$ nm, $\epsilon_F = 0.15$ eV. Numbers on the curve are DC magnetic field $B_0$, excitation at port 2.
	}
	\label{figure17}
\end{figure}

	In this subsection, we show the influence of the gaps between the nanodisk and the waveguides on the central  frequency of the circulator. The gaps was varied from 2.5 nm to 10 nm and the other parameters were fixed, i.e. $\epsilon_F = 0.15$ eV, $R$ = 320 nm and $w$ = 120 nm. The frequency responses are plotted in Fig. \ref{figure14} where one can see that  the central frequency of operation of the device, the magnetic field and the bandwidth decrease with the increase of the gap (see Fig. \ref{figure15}, Fig. \ref{figure16} and Fig. \ref{figure17}, respectively).
	\subsection{Control by  Fermi Energy}
\begin{figure}[h!]
	\centerline{
		\includegraphics[scale=1]{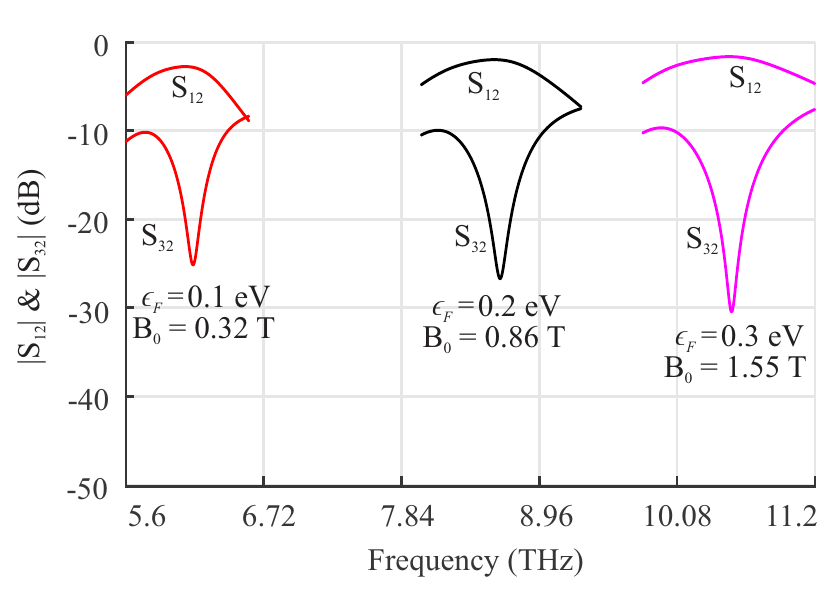}}
	\caption{Frequency responses of the circulator,  $g$ = 5 nm, $R$ = 320 nm, $w$ = 120 nm for different Fermi energies, excitation at port 2.
	}
	\label{figure18}
\end{figure}

\begin{figure}[h!]
	\centerline{
		\includegraphics[scale=1]{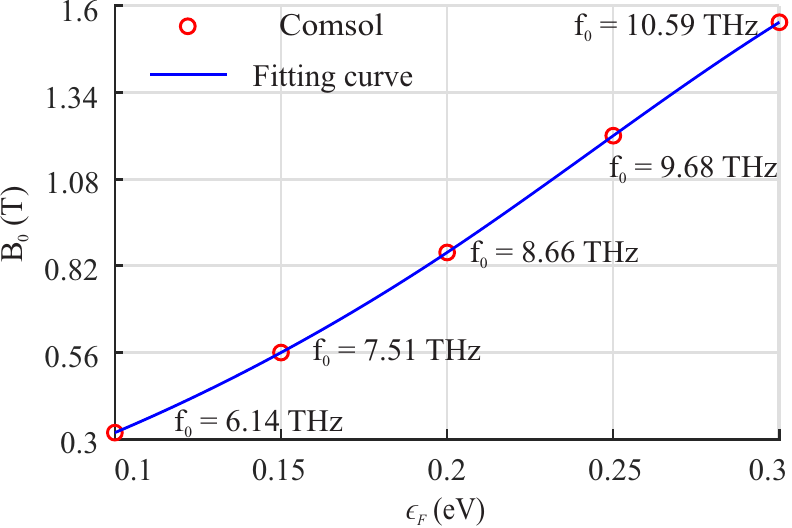}}
	\caption{Fermi energy dependence of the optimal magnetic field $B_0$ for circulator,  $g$ = 5 nm, $R$ = 320 nm, $w$ = 120 nm. Numbers on the curve give central frequency ${f}_{0}$, excitation at port 2.
	}
	\label{figure19}
\end{figure}

	To investigate the influence of the Fermi energy on the resonance frequencies, we varied $\epsilon_{F}$ from 0.1 eV to 0.3 eV.  The frequency responses of the circulator are plotted in Fig. \ref{figure18} from which it can be seen that the increase of the Fermi energy shifts the resonance frequency to higher values. Fig. \ref{figure19} shows increase almost linearly of the optimal
magnetic field $B_0$ with the increase of the Fermi energy. This is consistent with the formula   $\omega_B = {eB_0v_{F}^{2}}/{\epsilon_F}$. 	

\begin{figure}[h!]
	\centerline{
		\includegraphics[scale=1]{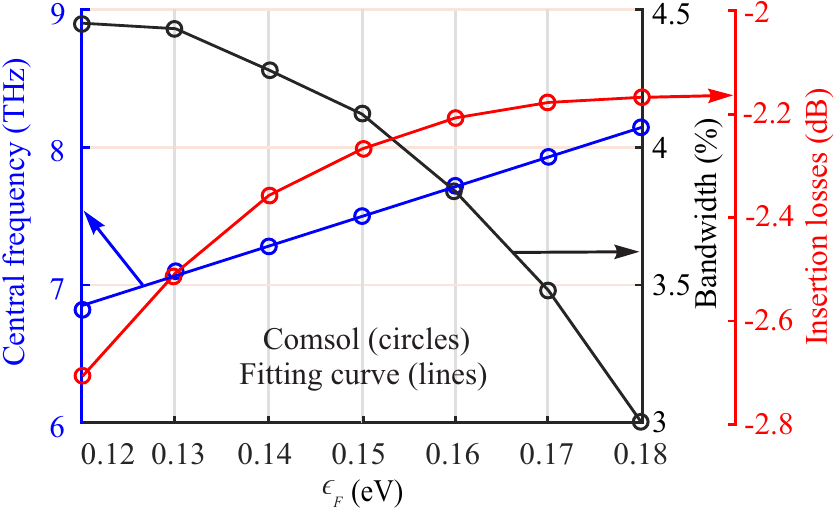}}
	\caption{Illustration of dynamical control by Fermi energy: central frequency (blue circle), bandwidth (black circle) and of insertion losses (red circle), $B_{0} = 0.56$ T, excitation at port 2.
	}
	\label{figure20}
\end{figure}

	To show a possibility  to control dynamically the central frequency of the device, we keep all the parameters of the circulator fixed and change only the Fermi energy.	Fig. \ref{figure20} shows the shift of the central frequency of operation to the higher frequencies with increase of $\epsilon_{F}$ and this is consistent with (\ref{eq:sig4}). With a small change of $\epsilon_{F}$ between 0.12 eV and 0.18 eV, the central frequency varies from 6.8 THz to 8.1 THz ( $\pm 10$\%).  This leads also to small changes in the insertion  losses and in the bandwidth of the device. Notice that in this case the DC magnetic field was fixed at $B_0 = 0.56$ T.
	\section{CONCLUSIONS}
	We have suggested and confirmed by  numerical simulations a realizability of  controllable three-port  graphene-based W-circulator operating in THz region.  The temporal coupled mode theory was  used for analysis of the circulator. The circulator presents a very simple structure consisting of graphene elements placed on a two-layer dielectric substrate. The central frequency of the circulator is defined mostly   by the radius of the disc resonator. This frequency depends also on the graphene waveguide width, the gap between the resonator and the waveguides, the DC magnetic field and on the Fermi energy of graphene. The device simulations demonstrated that at the central frequency of 7.5 THz, its bandwidth is  4.25\% with respect to the level of isolation -15 dB, and the insertion losses around -2 dB with the bias DC magnetic field $B_0 = 0.56$T and the Fermi energy of graphene  $\epsilon_F=0.15$eV. We have shown that the central frequency of the circulator can be dynamically shifted by $\pm 10$\% by an external electric field without significant deterioration of the circulator characteristics.  The presented results can be useful in the projects of THz graphene circuits.
		
	\section*{ACKNOWLEDGMENT}
	This study was financed in part by the Brazilian National Council for Scientific and Technological Development (CNPq).




\end{document}